\documentclass[11pt,superscriptaddress,aps,prd,preprint]{revtex4}
\usepackage{amsmath}

\makeatletter
\usepackage[T1]{fontenc}
\usepackage{amsmath}
\usepackage{amssymb}
\usepackage{graphicx}
\usepackage{tabularx}

\usepackage{slashed}

\usepackage{xcolor}

\newcommand{\bea}{\begin{eqnarray}}
\newcommand{\eea}{\end{eqnarray}}

\newcommand{\e}{\epsilon}

\begin{document}

\title{Lorentz violation, M\"{o}ller scattering and finite temperature}

\author{Alesandro F. Santos}\email[]{alesandroferreira@fisica.ufmt.br}
\affiliation{Instituto de F\'{\i}sica, Universidade Federal de Mato Grosso,\\
78060-900, Cuiab\'{a}, Mato Grosso, Brazil}

\author{Faqir C. Khanna\footnote{Professor Emeritus - Physics Department, Theoretical Physics Institute, University of Alberta\\
Edmonton, Alberta, Canada}}\email[]{khannaf@uvic.ca}
\affiliation{Department of Physics and Astronomy, University of Victoria,\\
3800 Finnerty Road Victoria, BC, Canada}

\begin{abstract}
Lorentz and CPT symmetries may be violated in new physics that emerges at very high energy scale, i.e., at the Planck scale. The differential cross section of the M\"{o}ller scattering, due to Lorentz violation at finite temperature is calculated. Lorentz-violating effects emerge from an interaction vertex due to a CPT-odd nonminimal coupling in the covariant derivative. The finite temperature effects are determined using the Thermo Field Dynamics (TFD) formalism. 

\end{abstract}

\maketitle


\section{Introduction}

The Standard Model (SM) respects the Lorentz and CPT symmetries that are supported by many experiments. Although the SM has achieved a remarkable phenomenological success there are still unresolved issues. Such problems emerge in a number of different scenarios, most of them related to new physics at the Planck scale $\sim 10^{19}\, \mathrm{GeV}$. Small Lorentz and CPT violations emerge in theories that unify gravity with quantum mechanics such as string theory \cite{Samuel}. There are several other issues that lead to Lorentz violation such as, loop quantum gravity \cite{Gambini, Alfaro}, geometrical effects such as noncommutativity \cite{Carroll, Jackiw}, torsion \cite{Tasson}, nonmetricity \cite{Foster} among others. A theory that allows incorporation of all Lorentz-violating terms together with the SM and general relativity is the Standard Model Extension (SME) \cite{SME1, SME2}. The SME is an effective field theory that preserves the observer Lorentz symmetry, while the particle Lorentz symmetry is violated. This model is divided into two versions, a minimal extension which has operators with dimensions $d\leq 4$ and a nonminimal version of the SME associated with operators of higher dimensions. Numerous possibilities have been investigated \cite{Kost3, Kost4, Kost5}.

The SME framework is one way to investigate Lorentz and CPT symmetries violation. A different way is to modify the interaction vertex adding a new nonminimal coupling term into the covariant derivative. This new nonminimal coupling term may be CPT-odd or CPT-even. There are various applications for both terms \cite{Belich1, Belich2, Belich3, Belich4, Belich5, SP, Brito, Casana1, Casana2, Casana3, Casana4}. In this paper the CPT-odd term is chosen to calculate the Lorentz violation correction to the electron-electron scattering, known as M\"{o}ller scattering. Corrections to the electron-electron scattering due to Lorentz violation have been studied \cite{Fu}. In this work the Z-boson exchange contribution to the amplitude is considered in addition to the electromagnetic interaction. In \cite{Fu} the minimal version of the SME has been used. Here the nonminimal CPT-odd term is used to calculate Lorentz violation corrections in the electron-electron scattering. In additon the finite temperature effects are calculated. The Thermo Field Dynamics (TFD) formalism is used to introduce finite temperature.

TFD formalism is a thermal quantum field theory \cite{Umezawa1, Umezawa2, Umezawa22, Khanna1, Khanna2} where the statistical average of an arbitrary operator is interpreted as the expectation value in a thermal vacuum. The thermal vacuum, defined by $|0(\beta) \rangle$, describes a system in thermal equilibrium, where $\beta=\frac{1}{k_BT}$, with $T$ being the temperature and $k_B$ the Boltzmann constant. This formalism depends on the doubling of the original Fock space, composed of the original and a fictitious space (tilde space), using  Bogoliubov transformations. The original and tilde space are related by a mapping, tilde conjugation rules. The physical variables are described by nontilde operators. The Bogoliubov transformation is a rotation involving these two spaces. As a consequence the propagator is written in two parts: $T = 0$ and $T\neq 0$ components. 

This paper is organized as follows. In section II, a brief introduction to the TFD formalism is presented. In section III, the  Lorentz violating nonminimal coupling is considered. The transition amplitude at finite temperature is determined. The differential cross section for M\"{o}ller scattering with Lorentz-violating parameter at finite temperature is calculated. In section IV, some concluding remarks are presented.

\section{Thermo Field Dynamics - TFD}

Here an introduction to TFD formalism is presented. It is a real time formalism of quantum field theory at finite temperature where the thermal average of an observable is given by the vacuum expectation value in an extended Hilbert space. This is achieved by defining a thermal ground state $|0(\beta) \rangle$. Then the expectation value of an operator $A$ is given as $\langle A \rangle=\langle 0(\beta)| A|0(\beta) \rangle$. However two main ingredients are required to achieve this: (1) doubling of degrees of freedom in the Hilbert space and (2) the Bogoliubov transformation. The doubling is defined by the tilde ($^\thicksim$) conjugation rules, where the expanded space is $S_T=S\otimes \tilde{S}$, with $S$ being the standard Hilbert space and $\tilde{S}$ being the fictitious space. The mapping between the tilde $\tilde{A_i}$ and non-tilde $A_i$ operators is defined by the following tilde (or dual) conjugation rules:
\bea
(A_iA_j)^\thicksim & =& \tilde{A_i}\tilde{A_j}, \quad (\tilde{A_i})^\thicksim = -\xi A_i,\\
(A_i^\dagger)^\thicksim &=& \tilde{A_i}^\dagger, \quad (cA_i+A_j)^\thicksim = c^*\tilde{A_i}+\tilde{A_j},\nonumber
\eea
with $\xi = -1$ for bosons and $\xi = +1$ for fermions. The Bogoliubov transformation being a rotation in the tilde and nontilde variables, thus introducing thermal quantities.

For bosons, Bogoliubov transformation are
\bea
a_p&=&\mathsf{u}'(\beta) a_p(\beta) +\mathsf{v}'(\beta) \tilde{a}_p^{\dagger }(\beta), \\
a_p^\dagger&=&\mathsf{u}'(\beta)a_p^\dagger(\beta)+\mathsf{v}'(\beta) \tilde{a}_p(\beta),\\
\tilde{a}_p&=&\mathsf{u}'(\beta) \tilde{a}_p(\beta) +\mathsf{v}'(\beta) a_p^{\dagger}(\beta), \\
\tilde{a}_p^\dagger&=&\mathsf{u}'(\beta)\tilde{a}_p^\dagger(\beta)+\mathsf{v}'(\beta)a_p(\beta),
\eea
where $\mathsf{u}'(\beta) =\cosh \theta(\beta)$ and $\mathsf{v}'(\beta) =\sinh \theta(\beta)$, with $a_p^\dagger$ and $a_p$ being creation and annihilation operators respectively. Algebraic rules for thermal operators are
\bea
\left[a(k, \beta), a^\dagger(p, \beta)\right]&=&\delta^3(k-p),\nonumber\\
 \left[\tilde{a}(k, \beta), \tilde{a}^\dagger(p, \beta)\right]&=&\delta^3(k-p),\label{ComB}
\eea
and other commutation relations are null.

For fermions, the Bogoliubov transformations are
\bea
c_p&=&\mathsf{u}(\beta) c_p(\beta) +\mathsf{v}(\beta) \tilde{c}_p^{\dagger }(\beta), \label{f1}\\
c_p^\dagger&=&\mathsf{u}(\beta)c_p^\dagger(\beta)+\mathsf{v}(\beta) \tilde{c}_p(\beta),\label{f2}\\
\tilde{c}_p&=&\mathsf{u}(\beta) \tilde{c}_p(\beta) -\mathsf{v}(\beta) c_p^{\dagger}(\beta),\label{f3} \\
\tilde{c}_p^\dagger&=&\mathsf{u}(\beta)\tilde{c}_p^\dagger(\beta)-\mathsf{v}(\beta)c_p(\beta),\label{f4}
\eea
where $\mathsf{u}(\beta) =\cos \theta(\beta)$ and $\mathsf{v}(\beta) =\sin \theta(\beta)$, with $c_p^\dagger$ and $c_p$ being creation and annihilation operators respectively. Here algebraic rules are
\bea
\left\{c(k, \beta), c^\dagger(p, \beta)\right\}&=&\delta^3(k-p),\nonumber\\
 \left\{\tilde{c}(k, \beta), \tilde{c}^\dagger(p, \beta)\right\}&=&\delta^3(k-p),\label{ComF}
\eea
and other anti-commutation relations are null.

An important note, the propagator in TFD formalism is written in two parts: one describes the flat space-time contribution and the other displays the thermal effect.

\section{Lorentz-violating corrections to M\"{o}ller scattering at finite temperature}
 
Our interest is to calculate the differential cross section for the process, $e^-(p_1)e^-(p_2)\rightarrow e^-(p_3)e^-(p_4)$, at finite temperature. This process is represented in the FIG. 1.
\begin{figure}[h]
\includegraphics[scale=0.6]{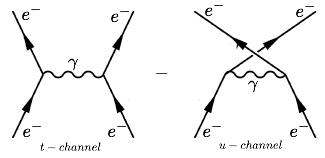}
\caption{M\"{o}ller Scattering}
\end{figure}

The lagrangian that describes the M\"{o}ller scattering is given by
\bea
{\cal L}&=& -\frac{1}{4}F_{\mu\nu}F^{\mu\nu}+\bar{\psi}\left(i\gamma^\mu D_\mu-m\right)\psi-\frac{1}{2\xi}\left(\partial_\mu A^\mu\right)^2.
\eea
To investigate the Lorentz violation in this scattering an alternative procedure is used. It consists of modifying just the SME interaction part via a non-minimal coupling using the covariant derivative, i.e.,
\bea
D_\mu=\partial_\mu+ieA_\mu+igb^\nu\tilde{F}_{\mu\nu},\label{D}
\eea
with $e$, $g$ and $b^\mu$ being the electron charge, a coupling constant and a constant four vector, respectively. $\tilde{F}_{\mu\nu}=\frac{1}{2}\epsilon_{\mu\nu\alpha\rho}F^{\alpha\rho}$ is the dual electromagnetic tensor with $\epsilon_{0123}=1$ and $F^{\alpha\rho}=\partial^\alpha A^\rho-\partial^\rho A^\alpha$. The new interaction breaks the Lorentz and CPT symmetries.  The interaction lagrangian is given by
\bea
{\cal L}_I=-e\bar{\psi}\gamma^\mu\psi A_\mu-gb^\nu\bar{\psi}\gamma^\mu\psi\partial^\alpha A^\rho\epsilon_{\mu\nu\alpha\rho}.\label{Int}
\eea
The second term represents the new interaction produced by the nonminimal coupling that leads to Lorentz violation, while the first term is the usual QED interaction. 
These vertices are represented as
\bea
\bullet\rightarrow V^\mu &=& -ie\gamma^\mu \\
\times \rightarrow gV_\rho &=& -gb^\nu\gamma^\mu q^\alpha \epsilon_{\mu\nu\alpha\rho},
\eea
where $q^\alpha$ is the momentum operator.  The diagrams in FIG. 1 represent the usual QED when the vertex $ V^\mu$ is used and the Lorentz violation corrections are represented by $  gV_\rho $ vertex. The total contribution is the sum of all diagrams.

In order to calculate the differential cross section for M\"{o}ller scattering the transition amplitude,
\bea
{\cal M}(\beta)=\langle f,\beta| \hat{S}^{(2)}| i,\beta\rangle,
\eea
is determined. Here $\hat{S}^{(2)}$ is the second order term of the $\hat{S}$-matrix that is defined as
\bea
\hat{S}&=&\sum_{n=0}^\infty\frac{(-i)^n}{n!}\int dx_1dx_2\cdots dx_n \mathbb{T} \left[ \hat{L}_{I}(x_1) \hat{L}_{I}(x_2)\cdots \hat{L}_{I}(x_n) \right],
\eea
where $\mathbb{T}$ is the time ordering operator and $\hat{L}_{I}(x)={L}_{I}(x)-\tilde{L}_{I}(x)$ describes the interaction. Here $L_I(x)$ and $\tilde{L}_I(x)$ are interaction lagrangian in usual and tilde Hilbert space respectively. The thermal states are
\bea
| i,\beta\rangle&=&c_{p_1}^\dagger(\beta)d_{p_2}^\dagger(\beta)|0(\beta)\rangle, \nonumber\\
 | f,\beta\rangle&=&c_{p_3}(\beta)d_{p_4}(\beta)|0(\beta)\rangle ,
\eea
with $\left(c_{p_j}^\dagger(\beta), d_{p_j}^\dagger(\beta)\right)$ and $\left(c_{p_j}(\beta), d_{p_j}(\beta)\right)$ being the  creation and annihilation operators respectively. Then the transition amplitude becomes
\bea
{\cal M}(\beta)&=&\frac{(-i)^2}{2!}\int d^4x\,d^4y\langle f,\beta|({\cal L}_I{\cal L}_I-\tilde{\cal L}_I\tilde{\cal L}_I)| i,\beta\rangle,\nonumber\\
&=&\left({\cal M}_1(\beta)+{\cal M}_2(\beta)+{\cal M}_3(\beta)\right)-\left(\tilde{\cal M}_1(\beta)+\tilde{\cal M}_2(\beta)+\tilde{\cal M}_3(\beta)\right)\label{amplit}
\eea
where
\bea
{\cal M}_1(\beta)&=&-\frac{e^2}{2}\int d^4x\,d^4y\,\langle f,\beta|\bar{\psi}(x)\gamma^\mu\psi(x)\bar{\psi}(y)\gamma^\nu\psi(y)A_\mu(x)A_\nu(y)| i,\beta\rangle,\\
{\cal M}_2(\beta)&=&-egb^\nu\epsilon_{\mu\nu\sigma\rho}\int d^4x\,d^4y\langle f,\beta|\bar{\psi}(x)\gamma^\omega\psi(x)\bar{\psi}(y)\gamma^\mu\psi(y)A_\omega(x)\partial^\sigma A^\rho(y)| i,\beta\rangle,\\
{\cal M}_3(\beta)&=&-\frac{1}{2}g^2b^\nu b^\rho\epsilon_{\mu\nu\alpha\sigma}\epsilon_{\omega\rho\delta\gamma}\int d^4x\,d^4y\langle f,\beta|\bar{\psi}(x)\gamma^\mu\psi(x)\bar{\psi}(y)\gamma^\omega\psi(y)\partial^\alpha A^\sigma(x)\partial^\delta A^\gamma(y)| i,\beta\rangle.
\eea
An important note, the ${\cal M}_3(\beta)$ contribution will be ignored since it is proportional to second order of Lorentz-violating parameter. Let's consider to the lowest order in this parameter. There are similar equations for the transition amplitude that include tilde operators. The fermion field is written as
\bea
\psi(x)=\int dp\, N_p\left[c_p u(p)e^{-ipx}+d_p^\dagger v(p)e^{ipx}\right],\label{fer}
\eea
where $c_p$ and $d_p$ are annihilation operators for electrons and positrons, respectively, $N_p$ is the normalization constant while $u(p)$ and $v(p)$ are Dirac spinors. There are two Feynman diagrams that describe the M\"{o}ller scattering, the $\mathbf{t}$-channel and the $\mathbf{u}$-channel, as represented in Fig. 1. For simplicity, 	consider
\bea
{\cal M}_1(\beta)={\cal M}_1^\mathbf{t}(\beta)+{\cal M}_1^\mathbf{u}(\beta),
\eea
where ${\cal M}_1^\mathbf{t}(\beta)$ and ${\cal M}_1^\mathbf{u}(\beta)$ are contributions due to the $\mathbf{t}$-channel and the $\mathbf{u}$-channel diagrams, respectively. Then using eq. (\ref{fer}) and Bogoliubov transformations we get
\bea
{\cal M}_1(\beta)&=&-\frac{e^2}{2} N_p\int d^4x\,d^4y\,\int d^4p(\mathsf{u}^2-\mathsf{v}^2)^2\bigl[\bar{u}(p_2)\gamma^\mu u(p_1)\bar{u}(p_4)\gamma^\nu u(p_3)e^{ix(p_1-p_2)}e^{iy(p_3-p_4)}\nonumber\\
&-& \bar{u}(p_2)\gamma^\mu u(p_3)\bar{u}(p_4)\gamma^\nu u(p_1)e^{ix(p_2-p_3)}e^{iy(p_4-p_1)}\bigl]\langle 0(\beta)|\mathbb{T}A_\mu(x)A_\nu(y)|0(\beta)\rangle.
\eea
The photon propagator at finite temperature is defined as 
\bea
\langle 0(\beta)|\mathbb{T}A_\mu(x)A_\nu(y)|0(\beta)\rangle=i\int \frac{d^4q}{(2\pi)^4}e^{-iq(x-y)}\Delta_{\mu\nu}(q,\beta),
\eea
with $\Delta_{\mu\nu}(q,\beta)\equiv\Delta_{\mu\nu}^{(0)}(q)+\Delta_{\mu\nu}^{(\beta)}(q)$, where $\Delta_{\mu\nu}^{(0)}(q)$ and $\Delta_{\mu\nu}^{(\beta)}(q)$ are zero and finite temperature components respectively. Explicitly these are
\bea
\Delta_{\mu\nu}^{(0)}(q)&=&\frac{\eta_{\mu\nu}}{q^2}\left( \begin{array}{cc}1 & 0 \\ 
0 & -1\end{array} \right),\\
\Delta_{\mu\nu}^{(\beta)}(q)&=&-\frac{2\pi i\delta(q^2)}{e^{\beta q_0}-1}\left( \begin{array}{cc}1&e^{\beta q_0/2}\\e^{\beta q_0/2}&1\end{array} \right)\eta_{\mu\nu}.\nonumber
\eea
Taking  $\mathsf{u}(\beta) =\cos \theta(\beta)$ and $\mathsf{v}(\beta) =\sin \theta(\beta)$ we get $(\mathsf{u}^2-\mathsf{v}^2)^2= \tanh^2(\frac{\beta |q_0|}{2})$, where $q_0=\omega$, and using the definition of the four-dimensional delta function,
\bea
\int d^4x\,d^4y\,e^{-ix(p_1-p_3+q)}e^{-iy(p_2-p_4-q)}=\delta^4(p_1-p_3+q)\delta^4(p_2-p_4-q),
\eea
the transition amplitude after carrying out the q integral becomes
\bea
{\cal M}_1(\beta)&=&-ie^2\Bigl[\bar{u}(p_2)\gamma^\mu u(p_1)\Delta^\prime(p_1-p_2,\beta)\bar{u}(p_4)\gamma_\mu u(p_3)\tanh^2\Bigl(\frac{\beta |(p_1-p_2)_0|}{2}\Bigl)\nonumber\\
&-& \bar{u}(p_2)\gamma^\nu u(p_3)\Delta^\prime(p_3-p_2,\beta)\bar{u}(p_4)\gamma_\nu u(p_1)\tanh^2\Bigl(\frac{\beta |(p_3-p_2)_0|}{2}\Bigl)\Bigl],
\eea
where $N_p=2$ has been used. The propagator is
\bea
\Delta^\prime_{\mu\nu}(q)\equiv \Delta^\prime(q)\,\eta_{\mu\nu}
\eea
with 
{\small
\bea
\Delta^\prime(q)=\frac{1}{q^2}\left( \begin{array}{cc}1 & 0 \\ 
0 & -1\end{array} \right)-\frac{2\pi i\delta(q^2)}{e^{\beta (q)_0}-1}\left( \begin{array}{cc}1&e^{\beta (q)_0/2}\\e^{\beta (q)_0/2}&1\end{array} \right).
\eea}
The remaining delta function, then has the overall four-momentum conservation, is taken out. In addition the center of mass is considered so that
\bea
p_1&=&(E,\vec{p}),\quad\quad p_2=(E,-\vec{p}),\nonumber\\
p_3&=&(E,\vec{p'})\quad\quad\mathrm{and}\quad\quad p_4=(E,-\vec{p'}),
\eea
where $|\vec{p}|^2=|\vec{p'}|^2=E^2$, $\vec{p}\cdot\vec{p'}=E^2\cos\theta$ and $s=(2E)^2=E_{CM}^2$, we get $|(p_1-p_2)_0|=|(p_3-p_2)_0|=E_{CM}$. Then 
\bea
{\cal M}_1(\beta)&=&-ie^2\Bigl[\bar{u}(p_2)\gamma^\mu u(p_1)\Delta^\prime(p_1-p_2,\beta)\bar{u}(p_4)\gamma_\mu u(p_3)\nonumber\\
&-& \bar{u}(p_2)\gamma^\nu u(p_3)\Delta^\prime(p_3-p_2,\beta)\bar{u}(p_4)\gamma_\nu u(p_1)\Bigl]\tanh^2\Bigl(\frac{\beta E_{CM}}{2}\Bigl).
\eea

In a similar way, the linear term in the Lorentz violating parameter, becomes
\bea
{\cal M}_2(\beta)&=&egb^\nu\epsilon_{\mu\nu\sigma\rho}\Bigl[(p_1-p_2)^\sigma\bar{u}(p_2)\gamma^\rho u(p_1)\bar{u}(p_4)\gamma^\mu u(p_3)\Delta^\prime(p_1-p_2,\beta)\nonumber\\
&-& (p_3-p_2)^\sigma\bar{u}(p_2)\gamma^\rho u(p_3)\bar{u}(p_4)\gamma^\mu u(p_1)\Delta^\prime(p_1+p_2)\Bigl]\tanh^2\Bigl(\frac{\beta E_{CM}}{2}\Bigl).
\eea

The differential cross section is given by
\bea
\left(\frac{d\sigma}{d\Omega}\right)_\beta=\frac{1}{64\pi^2 E_{\mathrm{CM}}^2}\cdot\frac{1}{4}\sum_{\mathrm{spins}}|{\cal M(\beta)}|^2. \label{CS}
\eea
Now an average over the spin of the incoming particles and sun over the spin of the outgoing particles is taken. Then the square transition amplitude is determined 
\bea
\bigl|{\cal M}(\beta)\bigl|^2=\bigl|{\cal M}_1(\beta)+{\cal M}_2(\beta)\bigl|^2.
\eea
In addition the relation
\bea
\bar{u}(p_2)\gamma_\alpha u(p_1)\bar{u}(p_1)\gamma^\alpha u(p_2)=\mathrm{tr}\left[\gamma_\alpha u(p_1)\bar{u}(p_1)\gamma^\alpha  u(p_2)\bar{u}(p_2)\right]
\eea
and the completeness relations:
\bea
\sum u(p_i)\bar{u}(p_i)&=&\slashed{p}_i+m
\eea
with $i=1, 2, 3, 4$ are used. Now the electron mass is ignored since the momenta are large, i.e., the ultrarelativistic limit ($p^2>>m^2$) is used.  

The differential cross section, eq. (\ref{CS}), at finite temperature for a time-like four vector $b^\nu=(b,0)$ becomes
\bea
\left(\frac{d\sigma}{d\Omega}\right)_\beta&=&\frac{1}{256\pi^2 s}\Biggl\{e^4\Biggl[\frac{-8\cos\theta+12\cos 2\theta+8\cos 3\theta+\cos 4\theta+115}{2(\cos\theta+1)^2}+C_1(\beta)\Biggl]\nonumber\\
&+&8e^2g^2E^4b^2\Bigl[\cos 2\theta+7+C_2(\beta)\Bigl]\Biggl\}\tanh^2\Bigl(\frac{\beta E_{CM}}{2}\Bigl),\label{eq40}
\eea
where
\bea
C_1(\beta)&\equiv&32E^4\Bigl\{\Delta^2(p_1-p_2)(\cos\theta+1)^2+4\left[\Delta(p_1-p_2)+\Delta(p_3-p_2)\right]^2\sin^4\theta/2\nonumber\\
&+&4\Delta^2(p_3-p_2)\Bigl\}\\
C_2(\beta)&\equiv&16E^4\Bigl\{\Delta^2(p_1-p_2)(\cos 2\theta+3)+4\Delta^2(p_3-p_2)\cos ^4\theta/2\Bigl\}
\eea
with
\bea
\Delta(q)=-\frac{2\pi i\delta(q^2)}{e^{\beta (q)_0}-1}\left( \begin{array}{cc}1&e^{\beta (q)_0/2}\\e^{\beta (q)_0/2}&1\end{array} \right).\label{eq43}
\eea

This result shows that corrections for M\"{o}ller scattering due to Lorentz violation at finite temperature are altered. Even if the Lorentz symmetry is conserved, there are corrections due to temperature for the usual result for electron-electron scattering. 

 An important note, in the Born approximation, the interaction potential may be calculated by using the Fourier transform of the scattering amplitude, i.e.,
\bea
V(r)=\frac{1}{(2\pi)^3}\int{\cal M}(\beta)\,e^{ip\cdot x}d^3p, 
\eea  
with ${\cal M}(\beta)$ being the total scattering amplitude given in eq. (20). In the non-relativistic limit some studies have been carried out about electron-electron interaction potential [30-32].  Our result, eq. (\ref{eq40}), has been obtained in the ultra-relativistic limit, i.e. $p^2>>m^2$. However using the non-relativistic limit (i.e., $p^2<<m^2$), in eq. (\ref{CS}), a different cross section at finite temperature and with Lorentz violating term is obtained. This result may be compared with the non-relativistic and Lorentz invariant differential cross section at zero temperature that is given as
\bea
\left(\frac{d\sigma}{d\Omega}\right)&=\frac{e^4}{4E^2}\left[(1-\cos^2\theta)+\frac{1}{1+\cos^2\theta}-\frac{1}{(1-\cos\theta)^2}\right],\label{NR}
\eea 
where $E$ is the total kinetic energy. In our case the potential is altered directly by Lorentz violation and temperature effects.

In addition the result obtained here may be compared with the non-relativistic result, eq. (45). Let's consider the following comparisons: (i) $b=0$ and $T\neq 0$ - 
In this case there are corrections due to temperature effects that become relevant in the high temperature limit since the estimates defined in eq. (43) become large. Then the dependent temperature part becomes dominant.  This result leads to a new motivation -  what are the modifications due to Lorentz violation at high temperature for electron-electron scattering? Then these estimates will give us a reasonable idea of the role of SME parameters at finite temperatures. (ii) $b\neq0$ and $T= 0$ - When the temperature is small the differential cross section is modified only by Lorentz-violating parameter.  Thus, while small, Lorentz violating terms do not contradict any experimental measurements of the M\"{o}ller scattering at zero temperature. However there is a motivation to perform such study at finite temperature. Although M\"{o}ller scattering at high energy has been investigated in experiments but these are still at zero temperature. There is certainly no investigation at extremely high energy with non-zero temperature that may indicate any role for Lorentz violation.

\section{Conclusion}

Lorentz and CPT symmetries are foundations of SM and General Relativity (GR). However violations of these symmetries could emerge in a new physics beyond the SM, at the Planck scale. The SME is an effective field theory that contain of SM, GR and all Lorentz- and CPT- violating operators. In parallel to SME exist an alternative procedure that consists in modifying the interaction part using a nonminimal coupling via covariant derivative. In this paper the nonminimal coupling term, CPT-odd term, is used to calculate the Lorentz-violating corrections to M\"{o}ller scattering at finite temperature. Finite temperature effects are introduced using the TFD formalism. Our results show that Lorentz-violating operators at finite temperature  contribute to the differential cross section of the electron-electron scattering. This is important  since there is no investigation at extremely high energy with non-zero temperature for this scattering. Here a theoretical study is developed such that relevant effects may arise for processes at very high energies. These results may modify the results that are anticipated for astrophysical processes. In addition, it is shown that the differential scattering cross section for M\"{o}ller scattering depends on temperature. The cross section also changes with Lorentz violation. Then this result may be useful to understand the role of Lorentz violation term depending on temperature. The interior of stars is a region where there is variation of the temperature, then stars may be used to test this result. At present this result is of theoretical interest and does not provide a direct way to measure an upper limit on the magnitude of the Lorentz-violating nonminimal coupling. Constraints on Lorentz-violating parameter can be obtained if measurements at finite temperature are realized in the future.  Although corrections due to squared Lorentz-violating parameters avoid the reach to stronger upper bounds, there are ways to constrain Lorentz violation. The dependence of the cross section on second order Lorentz- violating parameters has been obtained also in other scatterings. For  Bhabha scattering and for pair annihilation in the presence of a CPT-odd non-minimal coupling \cite{SP, Brito} and for electron-positron scattering with non-minimal CPT-even coupling term \cite{Casana1}. Good upper bounds for Lorentz-violating parameters have been obtained using experimental data from reference \cite{Derr}.

\section*{Acknowledgments}

This work by A. F. S. is supported by CNPq project 308611/2017-9.

\end{document}